\documentclass{imammb}
\jno{dqnxxx}
\usepackage{natbib}
\usepackage{graphicx}
\usepackage{epsfig}

\usepackage{amsmath,amsthm,bm,mathrsfs}

\numberwithin{equation}{section}

\begin{document}

\title{Delay effects in the response of low grade gliomas to radiotherapy: A mathematical model and its therapeutical implications}

\author{ {\sc V\'{\i}ctor M. P\'erez-Garc\'{\i}a$^*$}\\[2pt]
Departamento de Matem\'aticas, Universidad de Castilla-La Mancha \\
ETSI Industriales, Avda. Camilo Jos\'e Cela 3, 13071 Ciudad Real, Spain.\\
$^*${\rm Corresponding author: victor.perezgarcia@uclm.es} \\[6pt]
 {\sc Magdalena Bogdanska}\\[2pt]
Mathematics Department, Technical University of Lodz, \\
Lodz, Wolczanska 214 Street, Poland.\\[6pt]
{\sc Alicia Mart\'{\i}nez-Gonz\'alez, Juan Belmonte-Beitia}\\[2pt]
Departamento de Matem\'aticas, Universidad de Castilla-La Mancha \\
ETSI Industriales, Avda. Camilo Jos\'e Cela 3, 13071 Ciudad Real, Spain.\\[6pt]
{\sc Philippe Schucht} \\[2pt]
Universit\"atsklinik f\"ur Neurochirurgie \\
Bern University Hospital, CH-3010 Bern, Switzerland \\[6pt]
{\sc Luis A. P\'erez-Romasanta} \\[2pt]
Radiotherapy Unit, University Hospital of Salamanca, Salamanca, Spain \\[6pt]
{\rm [Received on  XXXXX]} \vspace*{6pt}}
\pagestyle{headings}
\markboth{ V. M. P\'EREZ-GARC\'{I}A \emph{et al.}}{\rm DELAY EFFECTS IN RADIOTHERAPY OF LGG}

\maketitle

\begin{abstract}
{Low grade gliomas (LGGs) are a group of primary brain tumors usually encountered in young patient populations. 
These tumors represent a difficult challenge because many patients survive a decade or more and may 
be at a higher risk for treatment-related complications. Specifically, radiation therapy is known to have a relevant effect on survival but in many cases it can be
 deferred to avoid side effects while maintaining its beneficial effect. However, a subset 
of low-grade gliomas manifests more aggressive clinical behavior and requires earlier intervention.
Moreover, the effectiveness of radiotherapy depends on the tumor characteristics. 
Recently Pallud et al., [Neuro-oncology, 14(4):1-10, 2012], studied patients with LGGs treated with radiation therapy as a first line therapy. 
and found the counterintuitive result that  tumors
with a fast response to the therapy had a worse prognosis than those responding late.
In this paper we construct a mathematical model describing the basic facts of glioma progression and response to radiotherapy. The model provides also an explanation to the observations of Pallud et al.  Using the model we propose radiation fractionation schemes that might be therapeutically useful by helping to evaluate the tumor malignancy while at the same time reducing the toxicity associated to the treatment.}
Low Grade Gliomas, Radiotherapy, Mathematical model of tumor response
\end{abstract}


\section{Introduction}

Low grade glioma (LGG) is a term used to describe WHO grade II primary brain tumors of astrocytic and/or oligodendroglial origin. 
These tumors are highly infiltrative and generally incurable but have median survival time of $> 5$ years because of low proliferation \citep{Pignatti2002,Pouratian2010}.
While most patients remain clinically asymptomatic besides seizures, the tumor transformation to aggressive high grade glioma is eventually seen in most patients. 

Management of LGG has historically been controversial because these patients are typically young, with few, if any, neurological symptoms. Historically, a wait and see approach was often favored in most cases of LGG due to the lack of symptoms in these mostly young and otherwise healthy adults. The support for this practice came from several retrospective studies showing no difference in outcome (survival, quality of life) if therapy was deferred \citep{Olson2000,Batchelor2006}. Other investigations have suggested a prolonged survival through surgery \citep{Smith2008}. In absence of a randomized controlled trial, recently published studies may provide the most convincing evidence in support of an early surgery strategy \citep{Jakola2012} and waiting for the use of other therapeutical options such as radiotherapy and chemotherapy. However, the decision on the individual treatment strategy is based on a number of factors including patient preference, age, performance status, and location of tumor \citep{Ruiz2009,Pouratian2010}.

As to radiation therapy the clinical trial by \citet{trial6} showed the advantage of using radiotherapy in addition to surgery. However, the timing of radiotherapy after biopsy or debulking is debated. It is now well known that immediate radiotherapy after surgery increases the time of response (progression-free survival), but does not seem to improve overal survival while at the same time inducing serious neurological deficits as a result to normal brain damage \citep{VandenBent2005}.
Overall survival depends more on patient and tumour-related characteristics such as age, tumour grade, histology and neurological function than the details of the plan of radiotherapy treatment.  
 Radioteraphy is usually offered for patients with a combination of poor risk factors such as age, sub-total resection, and diffuse astrocytoma pathology \citep{trial5}. 

Mathematical modelling has the potential to select patients that may benefit from early radiotherapy. Also it may help in developing specific optimal fractionation schemes for selected patient subgroups. However despite its enormous potential, mathematical modelling has had a limited use with strong focus on some aspects of radiation therapy (RT) for high-grade gliomas \citep{Rockne2010,BondiauRT,Konokoglu2010,Kirkby2010}. Up to now, no ideas coming from mathematical modelling have been found useful for clinical application.

There is thus a need for models accounting for the fundamental features of low-grade glioma dynamics and their response to radiation therapy without involving excessive details on the -often unknown- specific processes but allowing the qualitative understanding of the phenomena involved.  The availability of systematic and quantitative measurements of LGG growth rates provides key information for the development and validation of such models \citep{Pallud2012a,Pallud2012}.

In this paper we present a simple mathematical model capturing the key features seen in the response of LGGs to radiation. Our model incorporates the basic elements of tumor dynamics: infiltration and invasion of the normal brain by the tumor cells, proliferation and tumor cell death in response to therapy. Radiation therapy is included in an almost parameter-free way that captures the essentials of the dynamics and explains the relationship between proliferation, response to the therapy and prognosis as recently reported by \citet{Pallud2012}. 

In addition to explaining the counterintuitive observations of \citet{Pallud2012} the model presented in this paper can be used to explore different radiation regimes. The analysis to be presented in this paper suggests the possibility of using radiation therapy with palliative intent and also to test what the tumor response is and help the oncologists in making the best possible decisions on when and how to act on the tumor.

Our plan in this paper is as follows. First, in Section \ref{model} we present our model accounting for tumor cell dynamics and the response of the tumor cells to radiation. Next in Sec. \ref{results} we present the results of the numerical simulations of the model in different scenarios and study the dependence of the model on the different parameters. In Sec. \ref{analytical} we display some analytical estimates of the typical dynamics of the tumor response to radiation. In Sec. \ref{discussion} we discuss some hypothesis suggested from the model that may be useful for therapy. Finally, in Sec. \ref{conclusions} we summarize our conclusions.


\section{Mathematical model for the response of LGGs to radiotherapy}
\label{model}

\subsection{Tumor cell compartment}

In the last years there has been a lot of activity on mathematical models of glioma progression \citep{Murray2007,Stamatakos2006,Stamatakos2006b,Frieboes2007,Swanson2008,Bondiau2008,Eikenberry2009,Tanaka2009,Konokoglu2010,Wang2009,Rockne2010,PG2011,PG2012,Hatzikirou2012,Badoual2012,Alicia2012,Gu2012,Painter2013}. 
In this paper we will consider a model for the  compartment of tumor cells of the simplest possible type: a Fisher-Kolmogorov (FK) type equation \citep{Murray2007}. More complicated models such as 
 single-cell based models would allow, in principle, to follow the individual movement of the transformed astrocytes through the brain parenchyma. However, considering that the basic rules behind a model are more important than the model details, we have discarded both the use of on-lattice models, which are not realistic when cell motion is considered, and off-lattice models which assume fixed cell geometries and/or incorporate unknown cell-cell interactions. Besides, these models often require the estimation of a large number of unknown parameters and the determination of initial cell configurations, which are extremely difficult to validate in \emph{in vivo experiments} and/or using clinical data. Thus, to keep our description as simple as posible we have opted for a continuous model as follows
\begin{subequations}
\label{FK}
\begin{eqnarray}
\frac{\partial C}{\partial t} & = & D\Delta C+ \rho (1-C)C, \\
C(x,0) & = & C_0(x),
\end{eqnarray}
\end{subequations}

where $C(x,t)$ is the tumor cell density as a function of time $t$ and the spatial position $x$ and it is measured in units of the maximal cell density allowed in the tissue $C_*$ (typically around   $10^3$ cell/cm). $\Delta = \sum_{j=1}^n \partial^2/\partial x_j^2$ is the $n-$dimensional Laplacian operator. 

$D$ is the diffusion coefficient accounting for the average cellular motility measured in mm$^2$/day to be assumed in this paper to be constant and spatially uniform. Migration in gliomas is not simple and in fact many authors have proposed that the highly infiltrative nature of human gliomas recapitulates the migratory behavior of glial progenitors \citep{Suzuki2002,Dirks2001}. Here we assume, as in most models, that glioma cell invasion throughout the brain is basically governed by a standard Fickian diffusion process. More realistic and complicated diffusion terms in gliomas should probably be governed by fractional (anomalous) diffusion \citep{Fedotov2007} or other more elaborate terms \citep{Badoual2009} to account for the high infiltration observed in this type of tumors \citep{Onishi2011} and the fact that cells do not behave like purely random walkers and may actually remain immobile for long times before compelled to migrate to a more favorable place. In addition in real brain there are spatial inhomogeneities expected in the parameter values such as different propagation speeds in white and gray matter, and anisotropies (e.g. on the diffusion tensor with preferential propagation directions along white matter tracts).  Many papers have incorporated these details \citep{Benalli2005,Bondiau2008,Konokoglu2010,Clatz2005,Painter2013} mostly with the intention to make patient-specific progression predictions. However, the main limitation is the lack of information on the (many) patient specific unknown details what has limited progresses in that direction. Thus, in order to simplify the analysis and focus on the essentials of the phenomena we have chosen to study the model in one spatial dimension and in isotropic media.
It is interesting that up to now only the simplest models such as those given by Eqs. (\ref{FK}) have been used to extract conclusions useful for clinicians \citep{Swanson2008,Wang2009}. 

The choice of one-dimensional diffusion intends to incorporate qualitatively diffusion phenomena in the simplest possible way. Front solutions of the 1D FK equation have been extensively studied and are known to propagate with a minimal speed $c = 2\sqrt{\rho D}$ when starting from still initial data \citep{Murray2007}. It is interesting that the 1D model recapitulates the most relevant -for us-phenomenology observed in higher-dimensional scenarios. First it is obvious and well-known that front (invasion waves) solutions of the 1D FK equation also solve higher dimensional version of the equation \citep{Xin,Tyson}. Moreover, those solutions are asymptotically stable \citep{Xin,Sattinger} what means that unlike other more complicated non-symmetrical solutions \citep{Tyson}, they do arise as limits of non symmetric initial data. It is also well known that radially symmetric (in 2D) or spherically symmetric (in 3D) travelling wave solutions of FK do not exist in high dimensions but that symmetric fronts also develop in those scenarios with a non-constant speed that depends on the local curvature $R$ \citep{Tyson,Volpert}. As the front grows with time the now radius-dependent front speed is given by  $c(R) = c- D/R(t)$. Thus growing symmetric multidimensional solutions with large curvature radii ($R\rightarrow \infty$) grow with the same speed as 1D fronts \citep{Volpert,Gerlee}.

In this paper we are interested on the description of low-grade gliomas that typically are very extended when diagnosed, thus the initial data radius is large and fronts are well developed by then. Although during the initial stages of tumor development the dimensionality may play a relevant role, for spatially extended tumors the effect of using higher-dimensional operators is not expected to be substantial. 

Moreover, some of the phenomena to be described later in this paper are found to be essentially independent of diffusion and a very good qualitative agreement will be found between our simplified analysis and the growth dynamics of the mean tumor diameter. Taking into account all these evidences we will keep the system one-dimensional, since our intention is not to provide a detailed quantitative description of the processes -that in any way would be beyond the reach of a simple model such as FK- but instead to provide a qualitative description of the dynamics in the simplest possible way. As we will discuss in detail later, this approach will lead to a simple yet qualitatively correct description of the response of low grade gliomas to radiotherapy.

The parameter $\rho$ in Eq. (\ref{FK})  is the proliferation rate (1/day) its inverse giving an estimate of the typical cell doubling times. We have chosen a logistic type of proliferation leading to a maximum cell density $C(x,t) = 1$. Finally, the tumor evolves from an initial cell density given by the function $C_0(x)$ in an unbounded domain, so we implicitly assume it to be located initially sufficiently far from the grey matter. 

 A very interesting feature of model Eqs. (\ref{FK}) is the well known fact that a tumor front arises propagating at the asymptotic (constant) speed of $v_* = 2\sqrt{D\rho}$ what is in very good agreement with the observed fact that the tumor mean diameter grows at an approximately constant speed \citep{Pallud2012a}. 

While many other mathematical models of gliomas incorporate different cell phenotypes, e.g. normoxic (proliferative) and hypoxic (migratory) phenotypes, such as in \citet{Alicia2012,Hatzikirou2012,PG2012}, here we focus our attention on LGGs and as such will consider a single (dominant) tumor cell phenotype. In our model we do not include the possible existence of different tumor cell populations with different sensitivities to therapy such as stem cells since the function and mechanisms of stem cells in glioblastoma are yet under debate \citep{Barrett2012,Chen2012}.

\subsection{Response to radiation}

Radiation therapy has been incorporated in different forms to mathematical models of high-grade glioma progression \citep{Rockne2010,BondiauRT,Konokoglu2010,Kirkby2010}. In this paper we want to focus our attention on
 LGGs whose response to radiation is very different to the one observed in HGGs. Radiation therapy in LGGs 
 typically induces a response that prolongs for several years after therapy.

  Very interesting quantitative data on the response of LGGs to radiation have been reported in a retrospective study  by \citet{Pallud2012a}. The authors studied patients diagnosed with grade II LGGs treated with first-line radiotherapy before evidences of malignant transformation. 
Patients with a post-RT velocity of diametric expansion (VDE) \citep{Pallud2012a}  slower than -10 mm/year were taken as a subgroup of slowly growing LGGs. Patients with a post-RT VDE of -10mm/year or faster were included in the group of fast growing LGGs.
The authors concluded that the post-RT VDE was significantly faster in the group with high proliferation. Also in the patients with an available pre-RT VDE, the low pre-RT VDE subgroup presented a slower VDE at imaging progression. 
As to the survival time (ST), the post-radiotherapy VDE carried a prognostic significance on survival time, as the fast post-radiotherapy tumor volume decrease (VDE at -10 mm/year or faster) were associated with a significantly shorter survival  than the slow post-radiotherapy tumor volume decrease (VDE slower than -10 mm/year). 

The very slow response to radiotherapy, leading to tumor regression lasting for several years is difficult to understand in the context of the standard  linear quadratic model in which damage is instantaneous and leads to cell death early after radiation therapy. However a key aspect of the cell response to radiation is that irradiated cells usually die because of the so-called mitotic catastrophe after completing one or several mitoses \citep{Joiner2009}. This means that slowly proliferating tumors, as it is the case of LGGs with typical low proliferation indexes between 1\% and 5\% in pathological analyses need a very long time to manifest the accumulated cell damage that cannot be repaired. 

Thus, in order to capture in a minimal way the response of the tumor to radiation we will complement equation \eqref{FK} for the density of functionally alive tumor cells $C(x,t)$ with an equation for the evolution of the density of irreversibly damaged cells after irradiation $C_d(x,t)$. Our model for the evolution of both tumor cell densities will be given by the equations
\begin{subequations}
\label{modelocompleto}
\begin{eqnarray}
\frac{\partial C}{\partial t} &= & D\Delta C+ \rho (1-C-C_{d})C, \\
\frac{\partial C_{d}}{\partial t} & = & D\Delta C_{d} - \frac{\rho}{k} (1-C-C_{d})C_d. \label{damaged}
\end{eqnarray}
\end{subequations}
The first equation is a Fisher-Kolomogorov type equation describing the evolution of tumor cells $C(x,t)$. The saturation term includes the total tumor cell density, i.e. both the functional tumor cells and those damaged by radiation $C_d(x,t)$. The evolution of cells irreversibly damaged by radiation is given by Eq. (\ref{damaged}). As it is well described in the literature \citep{Joiner2009}, most of these cells behave normally until a certain number of mitosis cycles, thus we will consider that after an average of $k$ mitosis cycles these cells die resulting in a negative proliferation. The longer survival time $k\tau$ with $\tau = 1/\rho$ being the tumor population doubling time, results in a reduced proliferation potential $\rho/k$, that is the coefficient used for the negative proliferation term. Thus, the parameter $k$ in Eq. (\ref{damaged}) has the meaning of the average number of mitosis cycles that damaged cells are able to complete before dying. As with the normal population the number of cells entering mitosis depends in a nonlinear way on both tumor cell populations (cf. last term in Eq.\eqref{damaged}).

We will assume a series of radiation doses $d_1, d_2, ..., d_n $ given at times $t_1, t_2, ..., t_n$. 
The initial data for the first subinterval will be given by the equations
\begin{subequations}
\label{condiciones-empalme}
\begin{eqnarray}
C(x,t_0) & = & C_0(x),\\
C_d(x,t_0) & = & 0.
\end{eqnarray}
The evolution of the tumor follows then Eqs. (\ref{modelocompleto}) until the first radiation dose $d_1$, given at time $t_1$. The irradiation 
results in a fraction of the cells (surviving fraction) able to repair the radiation-induced damage given by $S_f(d_1)$ and a fraction $1-S_f(d_1)$ of cells
unable to repair the accumulated damage thus feeding the compartment of irreversibly damaged cells. The subsequent evolution of the populations is again given by Eqs. (\ref{modelocompleto}) until the next RT dose is given. In general after each irradiation event we get 
\begin{eqnarray}
C(x,t^+_j) & = &  S_f(d_j) C(x,t_j^-),\\
C_d(x,t_j^+) &=& C_d(x,t_j^-) + \left[1-S_f(d_j)\right] C(x,t_j^-),
\end{eqnarray}
\end{subequations}
where $S_f(d_j)$ is the survival fraction after a dose of radiation $d_j$, i.e. the fraction of damaged tumor cells after irradiation that are not able to repair lethal damage and are incorporated to the compartment of damaged cells. For the doses to be considered independent the interval between doses (typically 1 day) has to be larger than the typical damage repair times (of the order of hours). The evolution of both cell densities between irradiation events is given by the PDEs (\ref{modelocompleto}).


\subsection{Parameter estimation}

Eqs. (\ref{modelocompleto}) together with the initial conditions for each subproblem (\ref{condiciones-empalme}) define completely the dynamics of a low grade glioma in the framework of our simplified theoretical approach. 

We have chosen the parameters to describe the typical growth patterns of LGG. For the proliferation rate we have chosen typical values to be small and around $\rho=0.003$ day$^{-1}$ (see e.g. \citet{Badoual2012}), that give doubling times of the order of a year. Specifically we have considered values ranging from $\rho = 0.001$ day$^{-1}$  for very slowly growing LGGs to $\rho = 0.01$ day$^{-1}$.
For the cell diffusion coefficient we have taken values around $D=0.0075 \text{ mm}^{2}$ /day \citep{Benalli2005}. This choice, together with the previously chosen $\rho$ leads to asymptotic tumor diameter growth speeds  given by $v=4\sqrt{D\rho}$
of the order of several millimeters per year, in agreement with typical diametric growth speeds of LGGs \citep{Pallud2012a}. However the fact that the asymptotic speed is only reached when the tumor cell density is around one may require taking larger values of $D$ to match the real growth speeds.

As to the radiobiological parameters, being gliomas very resistant to radiation we have taken values in the range $S_f(\text{1.8 Gy}) \equiv \text{SF}_{1.8} \sim 0.9$ considering the median survival fraction value 0.83 after one dose of 2 Gy given by \citet{Kirkby2010} . 


Finally, the average number of mitosis cycles completed before the mitotic catastrophe occurs is difficult to estimate. This parameter intends to summarize in a single number a complex process in which a cell hit by radiation and its progeny die after some more mitosis cycles leading to a final extinction after a variable time. Death by mitotic catastrophe implies a minimal value of $k = 1$ and to allow for some more time we may choose values in the range $k=1-3$ \citep{Joiner2009}. We will show later that the choice of this parameter has a limited effect on the model dynamics and that in standard fractionation schemes there may be biological reasons to take it as $k=1$.

Our typical choices for the full set of model parameters is summarized in Table \ref{tableparameters}.

\begin{table}[h]
\begin{center}
\label{parameters}
\begin{tabular}{llll}
\hline\noalign{\smallskip}
Variable  & Description & Value (Units) & References \\
\noalign{\smallskip}\hline\noalign{\smallskip}
$C_*$& Maximum tumor  & $10^6$ cell/cm$^2$ & \citet{Swanson2008} \\ 
                  &     cell density      &                                       & \\ \noalign{\smallskip}\hline\noalign{\smallskip}
$D$& Diffusion coefficient for  & 0.01 mm$^{2}$ /day & \citet{Benalli2005}\\
          &     tumor cells                                                   &                                                   &    \\ \noalign{\smallskip}\hline\noalign{\smallskip}
$\rho$ & Proliferation  & 0.00356 day$^{-1}$ &  \citet{Badoual2012} \\
          &    rate                                                     &                                                  \\ \noalign{\smallskip}\hline\noalign{\smallskip}
SF$_{1.8}$ & Survival fraction & $\sim$ 0.9 & \citet{Kirkby2010} \\
         & for 1.8 Gy  & & \\ \noalign{\smallskip}\hline\noalign{\smallskip}
  $k$ & Average number of mitosis & 1-3 & \citet{Joiner2009} \\
          & cycles completed before the &      &   \\
          & mitotic catastrophe  & & \\        
\noalign{\smallskip}\hline\hline
\end{tabular}
\caption{Typical values of the biological parameters in the model of LGG evolution. \label{tableparameters}}
\end{center}
\end{table}
 
  In this paper we will fix the dose per fraction in agreement with the standard fractionation schemes for LGGs to be 1.8 Gy,  the only relevant parameter is the survival fraction SF$_{1.8}$ that will be taken to be around SF$_{1.8} \sim 0.9$, as discussed above.
In many examples we will choose the radiotherapy scheme as the standard fractionation of a total of 54 Gy in 30 fractions of 1.8 Gy over a time range of 6 weeks (5 sessions per week from monday to friday).  
 
\begin{figure}
\begin{center}
\epsfig{file=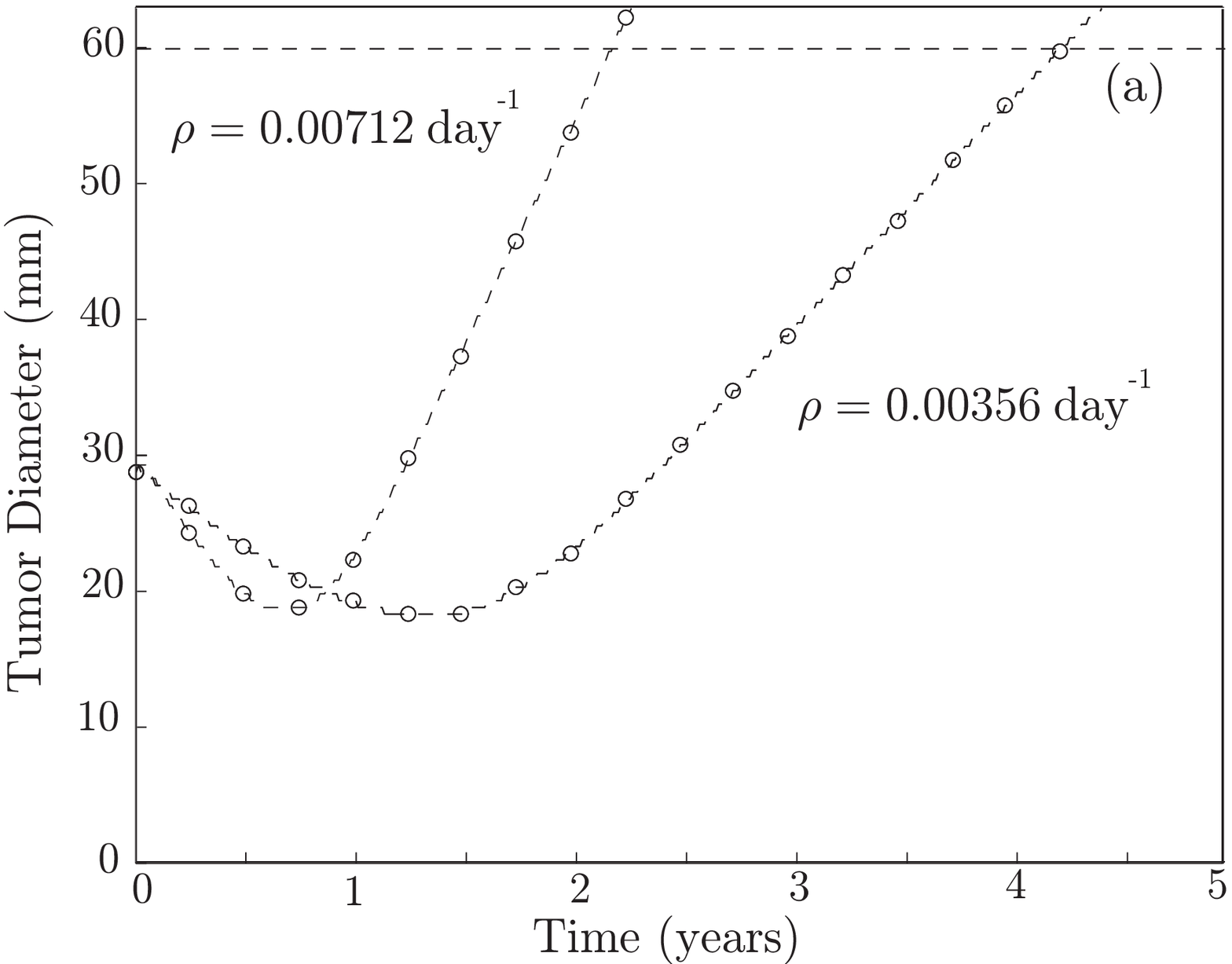,width=0.7\textwidth}

\ \ \ \ \ \ \ \  \ \epsfig{file=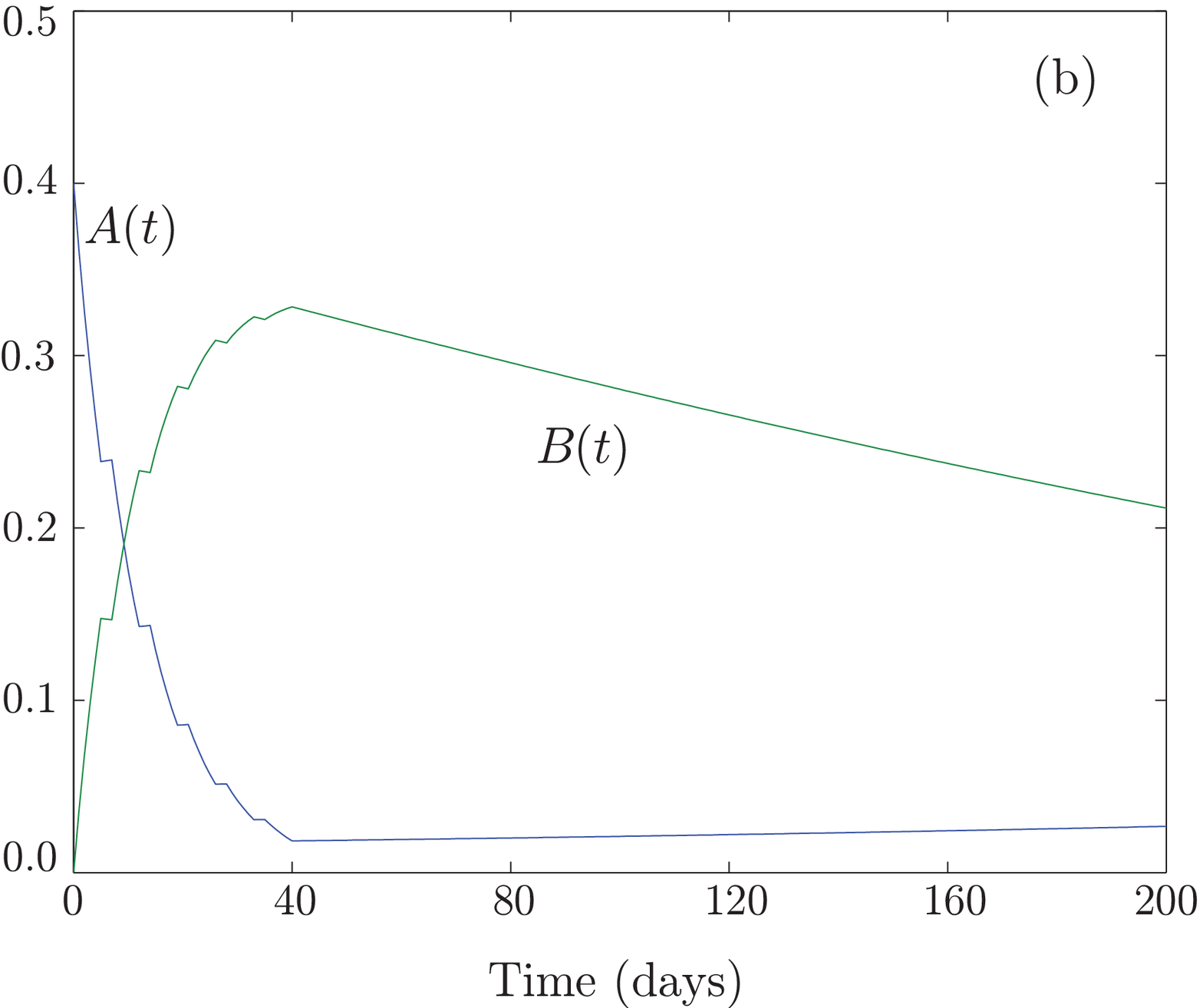,width=0.67\textwidth}
\end{center}

\caption{(a) Tumor diameter evolution for two different values of the proliferation:  $\rho = 0.00356$ day$^{-1}$, $\rho = 0.00712$ day$^{-1}$. Other parameter values are: $ D= 0.0075$ mm$^2$/day, critical detection value $C_{th} = 0.07$, SF$_{1.8} = 0.90$ and $k = 1$. The initial tumor cell densities are taken as $C_d(x,0) = 0, C(x,0) = 0.4 \text{sech}\left(x/6\right)$ with $x$ measured in mm, what gives an initial tumor diameter of 28.8 mm. Radiotherapy follows the standard scheme (6 weeks with 1.8 Gy doses from monday to friday) and starts at time $t=0$. Circles denote measurements every three months that would correspond to a close follow-up of the patient. The upper dashed line (horizontal) shows the fatal tumor burden size taken through this paper to be 6 cm as discussed in the text. 
(b) Evolution of the tumor cell amplitudes $A(t) = \max_x \left|C(x,t)\right|$ and  $B(t) = \max_x \left|C_d(x,t)\right|$ during the first 200 days showing the early response to the therapy for  $\rho = 0.00356$ day$^{-1}$.
\label{prima}}
\end{figure}

 \begin{figure}
\centering
\epsfig{file=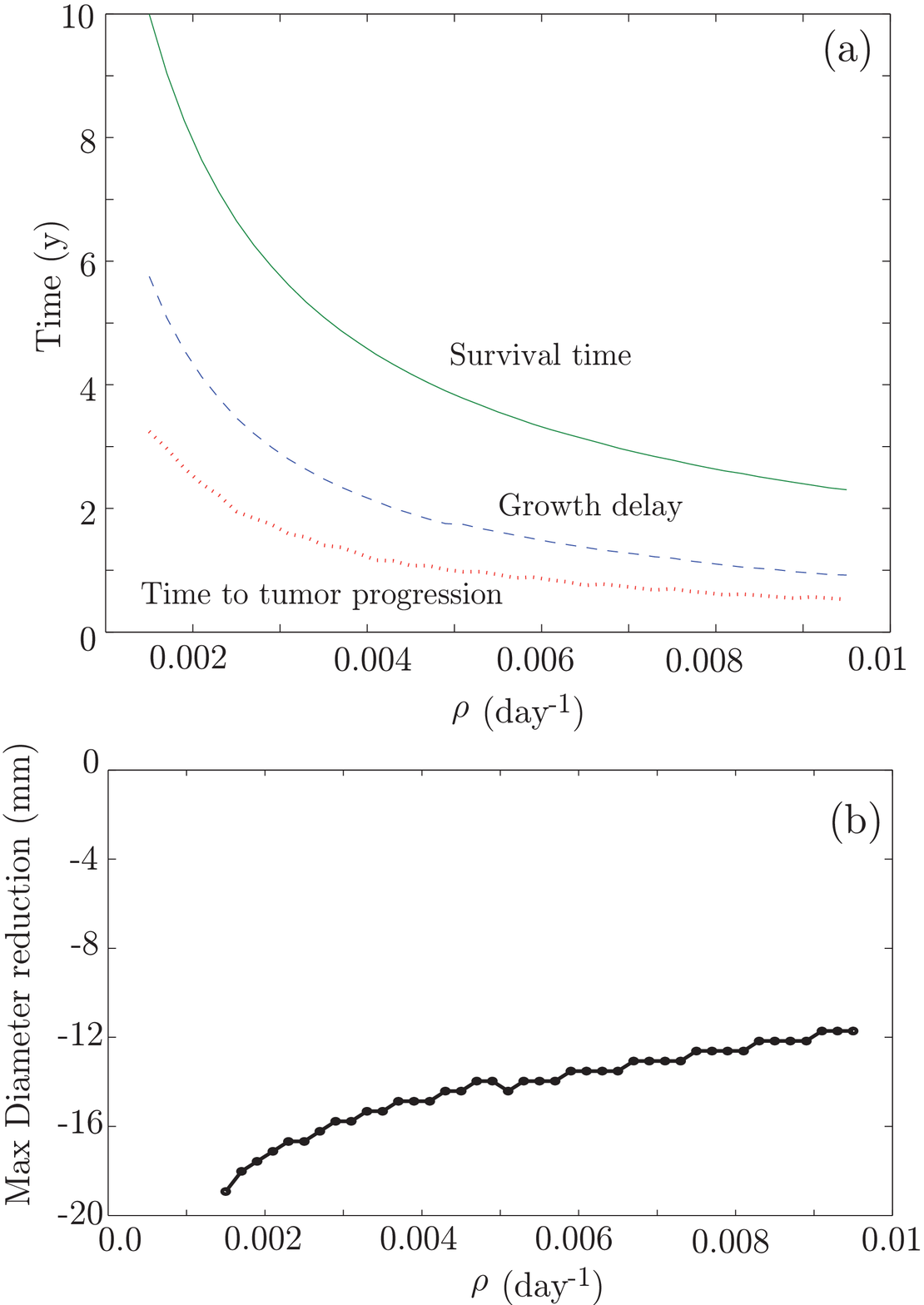,width=0.7\textwidth}
\caption{Dependence of the (a) survival time (solid line), growth delay (dashed line), time to tumor progression (dotted line) and (b) maximal reduction in diameter as a function of $\rho$. The curves summarize the outcome of many individual simulations with 
initial data
 $C_d(x,0) = 0, C(x,0) = 0.2 \exp\left(-x^4/81920\right)$ with $x$ measured in mm that gives an initial tumor diameter of 33.80 mm. Radiotherapy follows the standard scheme (6 weeks with 1.8 Gy doses from monday to friday) and starts at time $t=0$.
Other parameters used in the simulations are as in Fig. \ref{prima}. 
\label{variarrho}}
\end{figure}

\section{Results}  
\label{results}

\subsection{Computational details}

We have studied the evolution of the tumor diameter using our model equations (\ref{modelocompleto}) and (\ref{condiciones-empalme}). To solve the partial differential equations we have used standard second order finite differences both in time and space. Since the tumor diameter in the framework of this model tends to grow linearly in any spatial dimension we have focused on  the simplest one-dimensional version of the model. We have checked with simulations in higher dimensions that the dynamics is essentially the same and thus have sticked to the simplest possible model. To avoid boundary effects and focus on the  bulk dynamics, we have assumed our computational domain to be much larger than the tumor size.

In each simulation we have computed the tumor diameter as the distance between the points for which the density reaches a critical detection value $C_{th}$ that provides a signal in the T2 (or FLAIR) MRI sequence. Although which is that precise value is a debated question and in fact depends on the thresholds used in the images, we have assumed that $C_{th} \sim 0.05-0.07$. This is in agreement with the reported value of cellular density about 0.16 for detection \citep{Swanson2008} and a normal tissue density of about 0.1. In agreement with previous studies we take a fatal tumor burden (FTB) size of 6 cm in diameter \citep{Swanson2008,Wang2009}. As parameters containing useful information we have computed: the time in which the tumor starts regrowing after the therapy, usually called in clinical practice time to tumor progression (TTP), the time for which the tumor size equals its initial size -denoted as growth delay (GD)- and the time for which the tumor size equals the FTB or survival time (ST). 

We have studied a broad range of parameter values corresponding to the possible range of realistic values in the framework of our simple description of the tumor dynamics and its response to radiotherapy. We have also taken different types of initial data ranging from more localized (such as gaussian initial profiles) to more infiltrative (such as sech type functions). In what follows we summarize our results.

 \begin{figure}
\centering
\epsfig{file=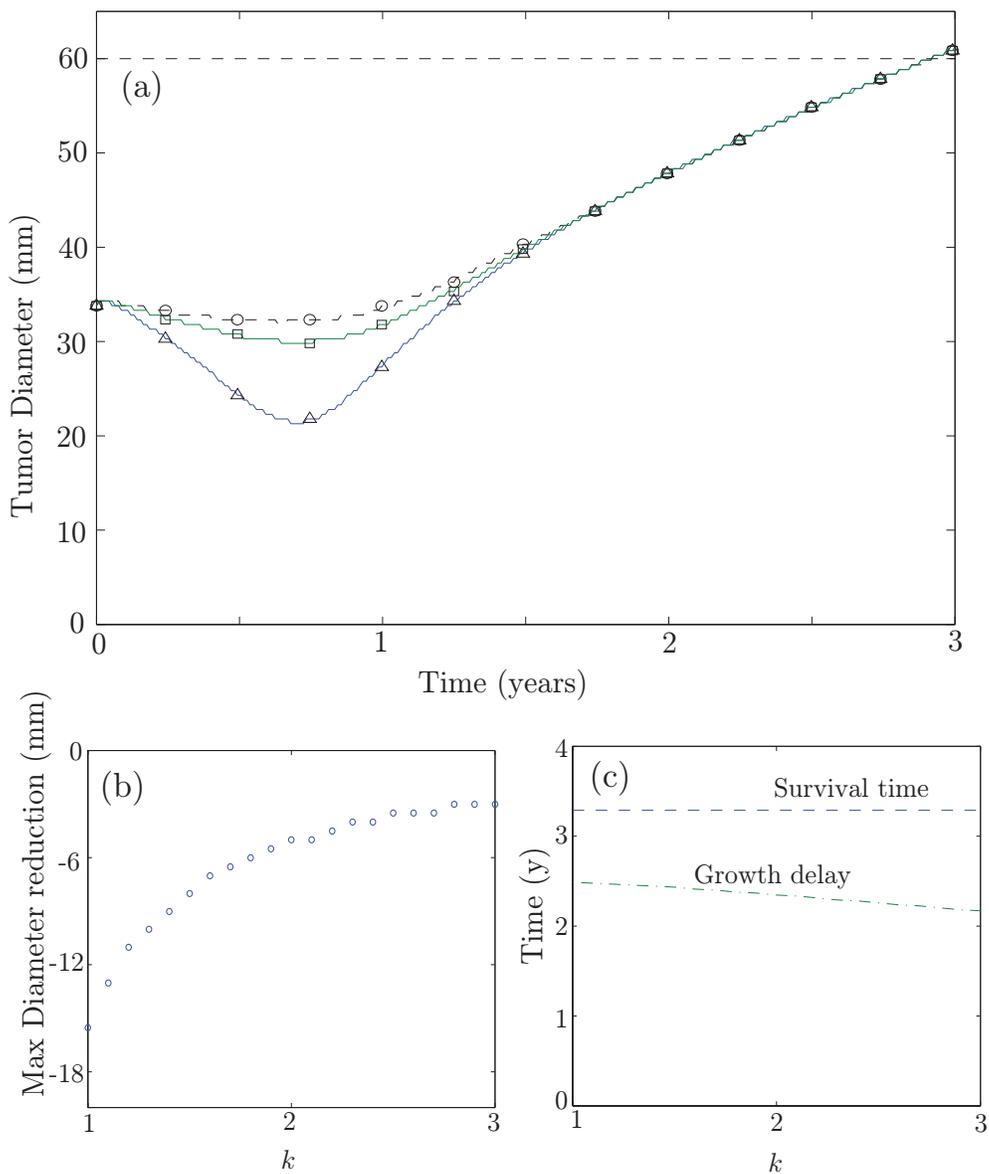,width=0.9\textwidth}
\caption{(a) Tumor diameter evolution for three different values of $k=1$  (triangles), $k=2$ (squares), and $k=3$ (circles) for $\rho = 0.00712$ day$^{-1}$. 
Other parameters are as in Fig. \ref{variarrho}. The upper dashed line (horizontal) corresponds to the fatal tumor burden size. 
Radiotherapy follows the standard scheme (6 weeks with 1.8 Gy doses from monday to friday) and starts at time $t=0$.
It is clear that for this set of parameters the survival time does not depend on $k$ despite the differences in the maximum diameter reduction achieved by the therapy. (b) Maximum diameter reduction for different values of the mean number of mitosis completed before cell death $k$ between 1 and 3. (c) Survival time and growth delay as a function of $k$. 
\label{segunda}}
\end{figure}

\subsection{Tumor proliferation rate determines the response to radiotherapy}

In a first series of simulations we have studied the dependence of the evolution of the tumor diameter on the proliferation rate. Fig. \ref{prima}(a) shows the evolution of the tumor diameter  for two different proliferation rates  $\rho = 0.00356$ day$^{-1}$ and 
$\rho = 0.00712$ day$^{-1}$ (Fig. \ref{prima}). In the first case of low proliferation, the tumor responds more slowly to therapy as measured by the speed of tumor regression (decrease in size) but the total response time is significantly longer, being the time to tumor progression of 16.9 months against 8.2 months in the later case. Also the growth delays are 14.7 months for the faster proliferating tumor against 29.9 months for the less proliferative one. Finally the ``virtual patient" with the slowly proliferating tumor survives much longer than that with the more proliferative one. This is just an example of a tendency shown in all of our simulations where \emph{more aggressive tumors respond earlier to the therapy}.  

It is remarkable that this fact is in full agreement with the results from \citet{Pallud2012}. Our model based on reasonable biological assumptions leads to a long remission time (e.g. in Fig. \ref{prima} of about 3 years) much larger than the treatment duration (6 weeks) and negatively correlated with the tumor proliferation rate. As a second relevant finding, also seen in the results shown by \citet{Pallud2012}, we observe that tumors responding faster have also shorter regrowth time. 

Fig \ref{prima}(b) shows the dynamics of the maximum density of tumor cells ($A(t) = \max_x C(x,t)$) and damaged tumor cells ($B(t) = \max_x C_d(x,t)$). As could be expected the amplitude of functionally alive tumor cells decreases during the therapy with the exception of the breaks in the weekends were a small increase is seen and correspondingly the amplitude of damaged tumor cells grows after every irradiation and for the full treatment period (6 weeks = 42 days). After $t=42$ days the population of tumor cells starts a slow recovery while the population of damaged cells declines in a much longer time scale. However, the width of the total tumor population evolves only in the slow time scale and does not display any effects during the treatment period.

It is important to emphasize that this behavior is not the result of a fortunate choice of the parameters but a generic behavior as we have confirmed through a large number of simulations covering the full clinically relevant parameter space. As an example, in  Fig. \ref{variarrho} we show how the variation of $\rho$ over a broad range of values leads to the same conclusion. Larger proliferation values accelerate the response but lead to earlier regrowth and as such, shorter growth delays and survival times [Fig. \ref{variarrho}(a)]. Our simulations also point out that the maximum volume reduction is only weakly dependent on the proliferation rate $\rho$ [Fig. \ref{variarrho}(b)], the smaller the proliferation rates the larger the maximum reduction in diameter.
 This fact is also in very good agreement with the results of \citet{Pallud2012} (see e. g. Fig. 2 bottom of their paper).

\subsection{The role of the number of mitosis before clonogenic cell death}

It is well known that most cells die after irradiation through the so-called mitotic catastrophe, i.e. due to incomplete mitosis, after completing one or several mitosis cycles. However, the specific choice of the parameter $k$ is not a priori obvious although a number between one and three is to be expected a priori from previous experience in vitro \citep{Joiner2009}. To get some information on how our model result's depend on this parameter we have explored numerically the range $k=1-3$. 

 Our results are summarized in Fig. \ref{segunda}. First, in Fig. \ref{segunda}(a) we show typical evolutions of the tumor diameter for three different values of $k=1$  (triangles), $k=2$ (squares), and $k=3$ (circles) for $\rho = 0.00712$ day$^{-1}$.  It is clear that although the diameter reduction depends on $k$ [Fig. \ref{segunda}(b)], the specific choice of this parameter does not affect the more relevant parameters such as the growth delay and the total survival and only weakly the growth delay [see Fig. \ref{segunda}(a),(c)]. From this and other simulations we think the specific choice of this parameter does not have a crucial role on the clinically relevant features dynamics. 

In addition, although in-vitro irradiation of cells with a single dose allows cells to complete a few mitosis cycles,  the accumulation of many doses in real treatment schedules implies that a typical cell receives a lot of DNA damage. This will probably make very difficult for cells in vivo to progress after the first mitosis, thus making reasonable to take $k=1$. This fact, together with the previous result on the independence of the clinically relevant endpoints on $k$ makes the choice of $k=1$ a reasonable assumption not expected to have a relevant impact on the final results.
 
\subsection{The role of cell motility}

\begin{figure}
\centering
\epsfig{file=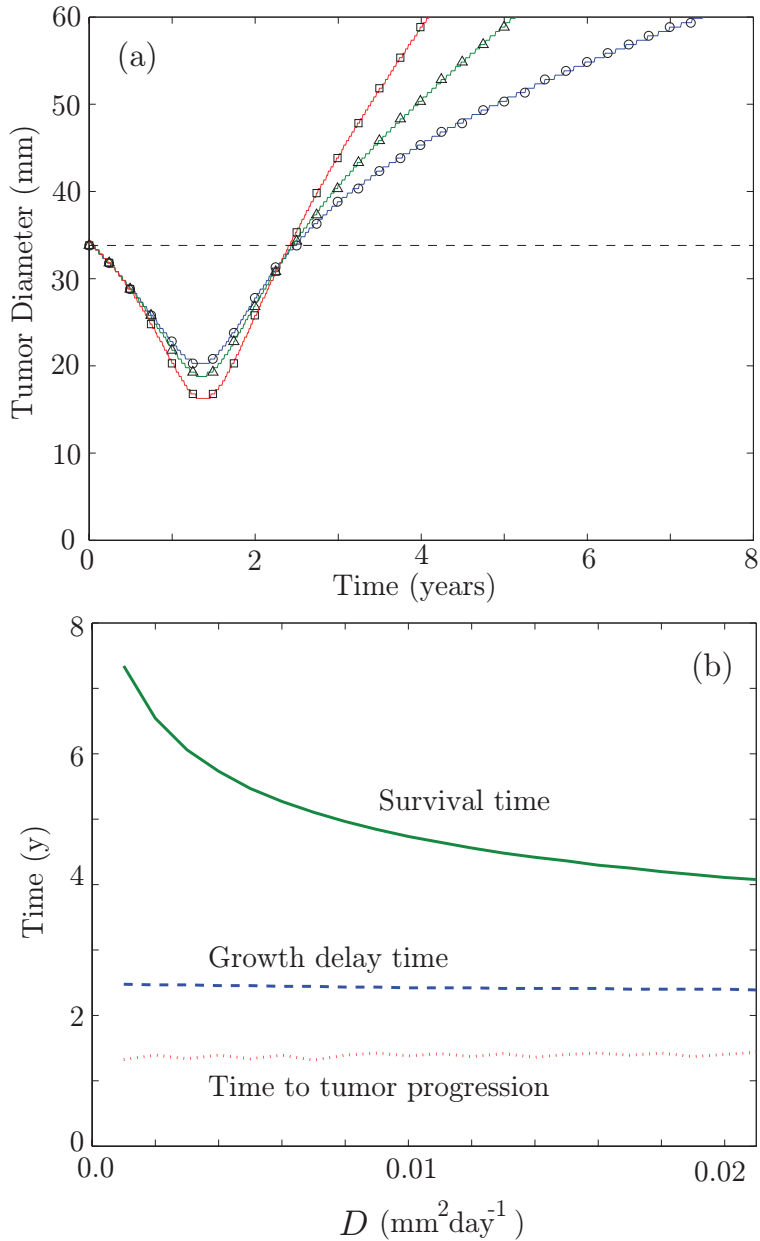,width=10cm}
\caption{(a) Tumor diameter evolution for three different values of $D=0.001$ mm$^2$day$^{-1}$ (circles), $D=0.007$ mm$^2$day$^{-1}$ (triangles), and $D = 0.021$ mm$^2$day$^{-1}$ (squares) for $\rho = 0.00356$ day$^{-1}$. Other parameters and initial data are as in Fig. \ref{variarrho}. Radiotherapy follows the standard scheme (6 weeks with 1.8 Gy doses from monday to friday) and starts at time $t=0$. It is clear that for this set of parameters the early response to the therapy does not depend on $D$ while the asymptotic growth does (b)  ST, GD and TTP as a function of $D$. Only the ST depends substantially on the cell motility $D$.
\label{figD}}
\end{figure}

We have also analyzed the role of the variation of the cell motility (invasion) parameter $D$. The typical outcome of several simulations for different values of this parameter is shown in  Fig. \ref{figD}. It is clear from Fig. \ref{figD}(a) that cell motility does not affect too much the dynamics of the response to the therapy except for long times because of the effect of the mobility on the asymptotic velocity of diametric expansion $v = 4 \sqrt{\rho D}$. This manifests in the independence of the time to tumor progression and growth delay on $D$  and the relevant impact of this parameter on the survival time (Fig \ref{figD}(b)).

\subsection{Deferring radiotherapy does not affect survival time}

\begin{figure}
\centering
\epsfig{file=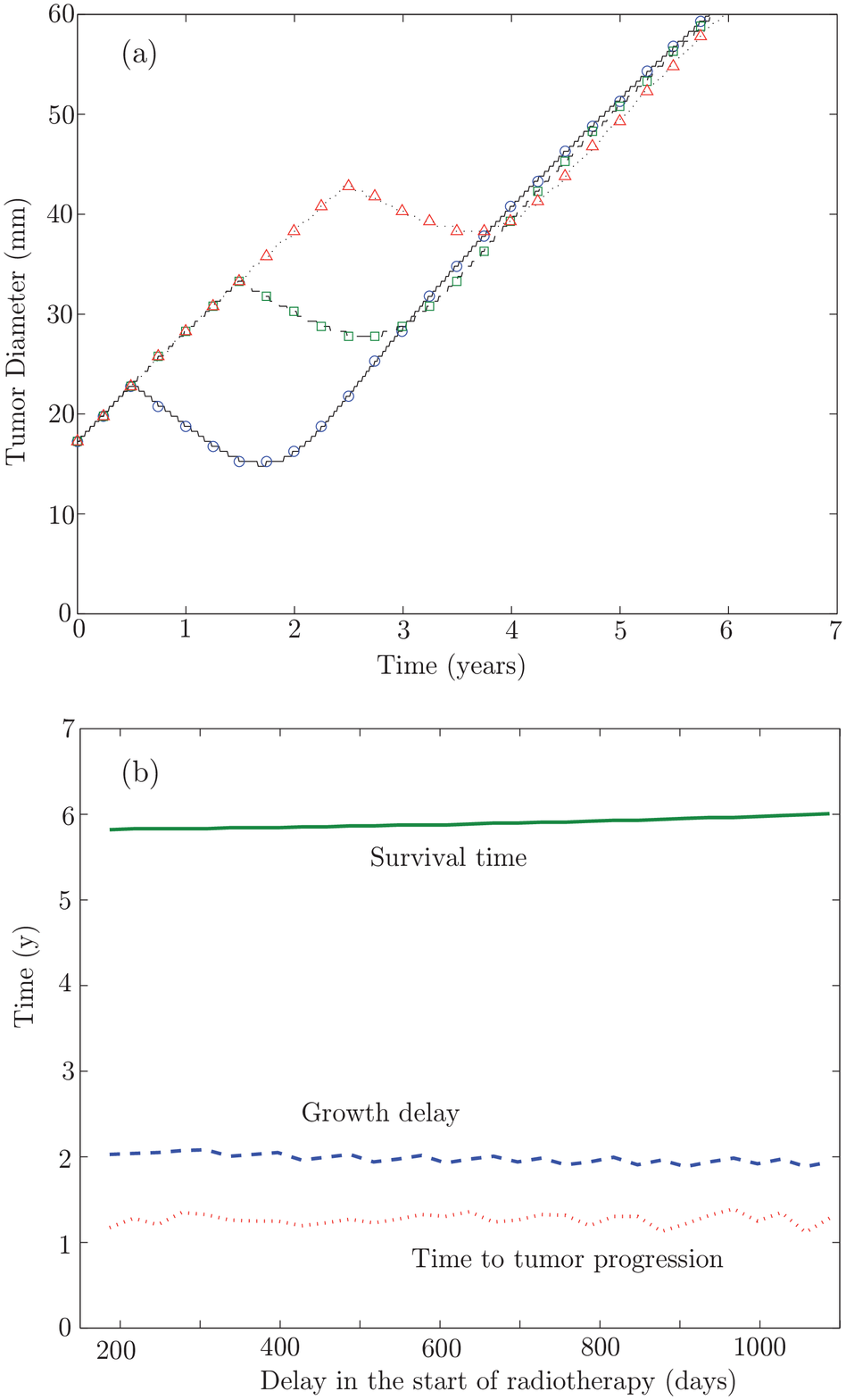,width=10cm}
\caption{Evolution of an initial tumor density given by  $ C(x,0) =0.2 \exp(-x^2/(2w^2)), C_d(x,0) = 0$ for $w = 6$, being ($x,w$) measured in mm, and parameter values
$D$ = 0.01 mm$^2$/day, $\rho = 0.004$ day$^{-1}$, SF$_{1.8}$ = 0.9 and $k=1$. Radiotherapy follows the standard scheme (6 weeks with 1.8 Gy doses from monday to friday) and starts at a given time $T_{\text{RT}}$ after the begining of the simulation for $t=0$. Shownare
(a) Tumor diameter evolution for three different values of $T_{\text{RT}}$ = 6 months (solid line, circles),  $T_{\text{RT}}$ = 18 months (dashed line, squares), 
$T_{\text{RT}}$ = 30 months (dotted line, triangles).
(b)  Survival time (solid line), growth delay (dashed line) and time to tumor progression (dotted line) as a function of the delay $T_{RT}$ in the start of the radiotherapy
\label{defered}}
\end{figure}

One clinical facts on radiotherapy of LGGs that has been proven in the last years  is that deferring radiotherapy has no significant impact on the survival time \citep{Bauman1999,VandenBent2005}. To test if our model presents this behavior we have run several series of simulations with different delays in the start of the radiotherapy  and compared the outcome. Typical results are shown in Fig. \ref{defered}. The results of the model fully agree with this very well known fact what gives us more confidence in the model's predictive power, despite its simplicity. 

\subsection{Splitting doses does not affect survival time}
\label{split}

\begin{figure}
\centering
\epsfig{file=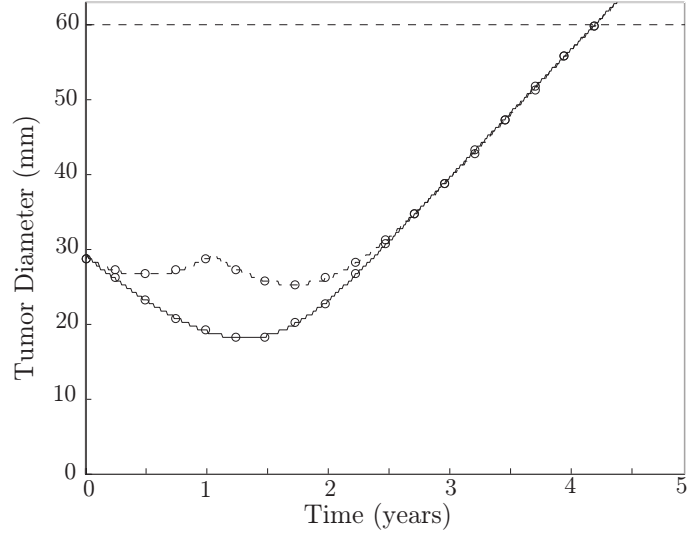,width=9cm}
\caption{Tumor diameter evolution for parameter values  $D=0.007$ mm$^2$/day, $\rho = 0.00356$ day$^{-1}$, initial data as in Fig. \ref{prima} and two different fractionations of the total dose. The solid line corresponds to the tumor evolution under 30 doses of 1.8 Gy given from monday to friday for six consecutive weeks starting 1 week after $t=0$. The dashed line corresponds to the tumor diameter evolution under the same fractionation scheme for the first three weeks and then deferring the remaining 15 doses for one year. Despite the time of response is shorter, the final survival time is the same in both fractionation schemes.
\label{figsplit}}
\end{figure}

We have studied the response of the tumor to radiotherapy under many different fractionation schemes maintaining the dose per fraction to be 1.8 Gy. Surprisingly all of the studied fractionations lead to very similar results for the virtual patient's survival time. A typical example is shown in Fig. \ref{figsplit}.

Although we have not tried every possible combination, this fact points out the difficultity of constructing specific fractionation schemes leading to a better outcome than those currently in use. However, the results of Fig. \ref{figsplit} have interesting potential practical applications as will be discussed in Sec. \ref{discussion}.

\section{Some analytical estimates}
\label{analytical}

The model equations given by Eqs. \eqref{modelocompleto}, though simple, do not have known analytical solutions allowing for the direct calculation of the clinically relevant quantities, i. e. the time to tumor progression ($t_{\text{TTP}}$), the growth delay ($t_{\text{GD}}$) and the time to fatal tumor burden. Even for the simplest version of the Fisher-Kolmogorov equations only a limited number of solutions are known for specific parameter values \citep{Murray2007,
Ablowitz1979}. 

Here we present some back-of-the-envelope calculations that may help in getting a qualitative idea of the typical dynamics of the tumor response to radiation. The basic idea behind our estimates is that during some time after irradiation the dominant component of the dynamics is the refilling of the compartment of the proliferating tumor cells and diffusion acts on a longer time scale being responsible for the asymptotic front speed (see e.g. Fig. \ref{figD}(a)) but having only a negligible influence both on the time to tumor progression and the growth delay (Fig. \ref{figD}(b)).

We will assume the tumor densities shortly after irradiation to be small enough to allow for the nonlinear terms to be neglected (it is in agreement to the low cellularity characteristic in LGG histologic samples). This is obviously true for the tumor compartment whose typical cell densities after irradiation are small until the tumor refills the space. As to the damaged tumor cell compartment its maximal density is of the order of the maximal initial tumor cell density (about 0.3-0.4) but decays in space to smaller cell densities and will be assumed to contribute only through the leading linear terms. 

\begin{figure}
\centering
\epsfig{file=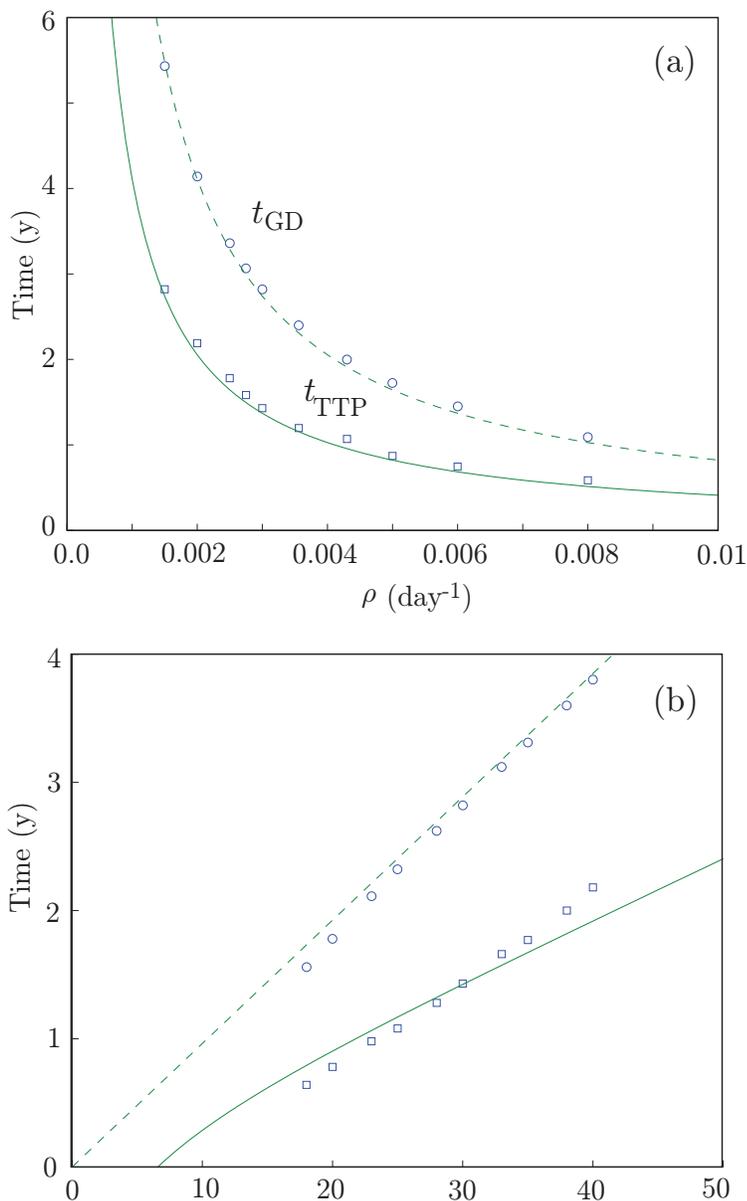,width=11cm}
\caption{Comparison of the estimates for $t_{\text{GD}}$ (circles) and $t_{\text{TTP}}$ (squares) obtained from Eqs. (\ref{tGD}) and (\ref{tTTP}) and their exact values (lines) obtained from 
numerical simulations of Eqs. (\ref{modelocompleto}) in different scenarios. In all cases $D=0.007$ mm$^2$/day, $S_f = 0.9$ and $k=1$ and initial data are as in Fig. \ref{prima}.
(a) Dependence on the proliferation parameter $\rho$ for a fixed number of doses $n=30$ following the standard fractionation scheme. (b) Dependence on the number of doses $n$ for fixed $\rho = 0.002$ day$^{-1}$. \label{validate}}
\end{figure}

As a final assumption, we will assume the total treatment time to be short in comparison with the typical proliferation times so that the effect of the radiotherapy will be incorporated through a modification of the pretreatment tumor cell density $C_0(x)$. Thus, for our rough estimates we will take 
\begin{subequations}
\begin{eqnarray}
C(x,0) & = &  S_f^n C_0(x), \\
C_d(x,0) & = &  \left(1-S_f^n\right) C_0(x).
\end{eqnarray}
\end{subequations}
Our set of hypothesis leads to a very simple evolution law for the total densities, valid for short times
\begin{subequations}
\begin{eqnarray}
C(x,t) & \approx &  S_f^n C_0(x) e^{\rho t},\\
C_d(x,t) & \approx &  \left(1-S_f^n\right) C_0(x) e^{-\rho t/k},
\end{eqnarray}
\end{subequations}
so that for some time after the therapy, the total tumor cell density $C_T(x,t)$ can be roughly approximated by 
\begin{equation}\label{aproxa}
C_T(x,t) \approx \left[ S_f^n e^{\rho t} +  \left(1-S_f^n\right) e^{-\rho t/k} \right] C_0(x),
\end{equation}
where $A(t) \equiv S_f^n e^{\rho t} +  \left(1-S_f^n\right) e^{-\rho t/k}$ provides an estimate of the tumor maximum density as a function of time. From this simple formula we can estimate the growth delay time since it would correspond to the time $t_{\text{GD}}>0$ such that 
\begin{equation}\label{alg}
S_f^n e^{\rho t_{\text{GD}}} +  \left(1-S_f^n\right) e^{-\rho t_{\text{GD}}/k} \approx 1.
\end{equation}
Although Eq. (\ref{alg}) is an algebraic equation with no simple explicit solutions by the time regrowth happens we can expect the first term to have a very small contribution while the second one would dominate what gives
\begin{equation} \label{tGD}
 t_{\text{GD}} \approx \frac{1}{\rho} \log \left(\frac{1}{S_f^n}\right) \approx \frac{n\left(1-S_f\right)}{\rho}
 \end{equation}
This equation incorporates the fact that the growth delay time does not depend much neither on the diffusion parameter $D$ (see e.g. Fig. \ref{figD}(b)), nor on the number of mitosis before cell death for damaged cells (see e.g. Fig. \ref{segunda}(c)) and points out a direct simple dependence of this time on the survival fraction, number of doses and proliferation parameter. Moreover the dependence of  $ t_{\text{GD}}$ on $\rho$ is  $\sim 1/\rho$ what resembles closely  the dependence depicted in Fig. \ref{variarrho}. 
We can also get estimates for the time to tumor progression since it corresponds to the point of minimum amplitude, corresponding to the time $t_{\text{TTP}}$ such that $A'(t_{\text{TTP}}) = 0$
\begin{equation}
\frac{d}{dt} \left[ S_f^n e^{\rho t} +  \left(1-S_f^n\right) e^{-\rho t/k }\right] = 0,
\end{equation}
what leads to 
\begin{equation} \label{tTTP}
t_\text{TTP} = \frac{1}{\rho\left(1+ \frac{1}{k}\right)} \log \left( \frac{1-S_f^n}{S_f^n}\right) \simeq \frac{n \left(1-S_f\right)}{\rho \left(1+ \frac{1}{k}\right)}
\end{equation}

While these estimates are obtained as rough approximations for the response to radiation, they provide a very reasonable agreement with the results of direct numerical simulations of Eqs. (\ref{modelocompleto}). For instance in Fig. \ref{validate}(a) we compare the results for the growth delay and time to tumor progression provided by Eqs. (\ref{tGD}) and (\ref{tTTP}) with the results from Eqs. (\ref{modelocompleto}) for a typical set of parameters and varying the proliferation parameter $\rho$. In Fig. \ref{validate}(b) we compare  the predictions for $t_{\text{TTP}}$ and $t_{\text{GD}}$ given by Eqs. (\ref{tGD}) and (\ref{tTTP}) for different values of the number of radiation fractions $n$, with the simulations of the full PDEs. 

In addition to these quantities it is possible to use Eq. (\ref{aproxa}) to get estimates for the conditions of response to therapy ($A'(0) <0$), i.e.
\begin{equation}
A'(0)= \rho S_f^n - \rho \left(1-S_f^n\right)/k < 0.
\end{equation}
This leads to the result that an estimate for the minimal number of doses leading to response is about 
\begin{equation} \label{minimaldose}
n > - \frac{\log\left(1+k\right)}{\log S_f}.
\end{equation}
Interestingly, this number is independent of $\rho$ and for the typical values of $S_f $ used in our simulations we get a minimal number of sessions around 7 and 9. 
 We have compared this estimate with the results of direct simulations of Eqs. (\ref{modelocompleto}) and found a very good agreement. For instance, taking typical initial data and parameter values $C(x,0) = 0.4 \text{sech}(x/11)$, $\rho = 0.00356 \ \text{day}^{-1}$, $k=1$, $D=0.0075$ mm$^2$/day and $S_f = 0.92$ we get response for $n \geq 10$ that is very close to the theoretical estimate computed from Eq. (\ref{minimaldose}) which is $n \geq 9$. The same happens for other parameter choices.

Finally, neither the growth delay as given by Eq. \eqref{tGD} nor the time to tumor progression \eqref{tTTP} depend on the initial amplitude or time $t_0$. This means that radiation can be deferred with no effect on these quantities, what matches very well the behavior observed in Fig. \ref{defered}(b).

\section{Discussion and therapeutical implications}
\label{discussion}

The intention of this paper is to propose a simple mathematical model adding radiation therapy in a minimal way to the simplest model of tumor progression. Despite its simplicity, the model reproduces many of the well-known facts of RT of LGGs as well as the recent results by \citet{Pallud2012}. The fact that the model reproduces so well what is known makes us wonder if it can be used to obtain any new information and/or to propose novel ideas with potential of translational application. In this section we make several proposals based on the mathematical model.

The first one is based on the fact, discussed in Sec. \ref{split}, that deferring part of the treatment does not affect survival time. This concept opens the door to dose fractionation approaches were part of the radiation is given right after surgery as an adjuvant therapy and the remaining radiation is given on progression (or later), thus controlling early the tumor while deferring in time the appearance of side effects. A specific way of implementing this idea would be to complete the radiation therapy exactly when the tumor size is minimal so that  the irradiated volume is substantially smaller than initially and the side effects due to so called volume effect would be reduced. Thus, instead of waiting till the tumor has extended substantially the first dose would be given right after resection and the second one on minimal volume. The fact that the tumor volume irradiated would be smaller might allow for the consideration of dose scalation protocols while maintaining the side effects under control.

A second implication of our model is that delivery of a reduced radiation dose larger than the minimal response dose (e.g. 30 Gy) should have a verifiable effect on tumor size following subtotal surgery up front as adjuvant therapy. Monitoring the response of the tumor to radiation one could get an idea of its malignancy due to the relation between response time and proliferation and correlate the finding with the proliferation index obtained through immunohistochemistry when available. The toxicity of this approach for the normal brain is low since 30 Gy is about the same level  the prescriptions for even full-brain irradiation under metastatic spreading, a dose that is very well tolerated by the brain.

Having an early estimate of the tumor aggressiveness is a potentially interesting information since in addition to making survival estimates (from the tumor parameters $\rho, D$) it 
would allow to discriminate tumors that are not as benign as initially thought. A first reason is that 
the tumor may have transformed to some degree into a more malignant one. This happens sometimes when the histological analysis is old, and by the time RT starts the tumour has become malignant. A different relevant source of error on diagnosis is sampling error if the histology was obtained through biopsy. In fact, biopsy is known to underestimate glioma grade in roughly 30\% of cases \citep{Muragaki} due to localized malignant transformation outside the biopsy location. Those tumors with high proliferation values obtained from the mathematical modeling should be expected to have early (radiological) malignant transformation and then MRI's should be taken in shorter intervals, as suggested e.g. by \citep{Pallud2012} for fast growing tumors. A final therapeutic option for those tumors with fast growth and/or expected early malignant transformation should be to consider another surgery (if feasible) or to start chemotherapy.

Thus, taking together our two ideas we could make specific recommendations:  for tumors with a high risk of malignant transformation, i.e. short regrowth time our suggestion would be 
to complete the full radiation dosing while for those growing slowly one could wait either until the malignant transformation or to the point were the tumor starts regrowing. 

\section{Conclusions}
\label{conclusions}

In conclusion, in this paper we have constructed a mathematical model combining the standard Fisher-Kolmogorov dynamics for tumor cells with a model for the response to radiation based on radiobiological facts. 
Our equations provide a theoretical link between proliferation and response to therapy that is one of the main results of this paper.
The model predicts that tumors with high proliferation will respond faster to RT than those with slower proliferation values. However this regression would only be transient and a regrowth is expected early in those tumors responding faster.  This fact, despite being somewhat counterintuitive has been confirmed in very recent retrospective studies by \citet{Pallud2012}. The model also displays the observed behavior that deferring RT does not affect survival time.

The equations allow to obtain analytical estimations for the growth delay time, time to tumor progression and conditions of response to therapy such as the minimal number of doses leading to response.

In addition to describing the known features of the response of LGGs to radiotherapy the model allows to get interesting predictions that may be amenable to further research. One of them is to follow a split-dose approach with a fraction of the total amount of radiation being given after surgery and the remaining on progression. This methodology would allow to get information on the tumor growth parameters what may lead to, estimates of the expected time to malignant transformation, survival, etc., while at the same time reducing toxicity.

We hope that our results will stimulate further colaborative studies directed to improve the quality of life of patients suffering this devastating disease.

\section*{Acknowledgements} This work has been supported by grant MTM2012-31073 (Ministerio de Econom\'{\i}a y Competitividad, Spain) and the James S. McDonnell Foundation (USA) through the 
21st Century Science Initiative in Mathematical \& Complex Systems Approaches for Brain Cancer-Pilot Award
 220020351.
 
 \appendix
 
\section{Study of the system without diffusion}
\label{simplified}

\subsection{Motivation and simplified model}

The main focus of the paper is the obtention of results on LGG progression that is related to the tumor size if the transition to malignancy is not taken into account. However it is interesting to note that a lot of information on the kinetic part of equations \eqref{modelocompleto} can be obtained. Thus in this appendix we will study the pair of ordinary differential equations
\begin{subequations}
\label{modeloEDOs}
\begin{eqnarray}
\frac{dA}{dt} &= & \rho (1-A-B)A, \label{nodamaged}\\
\frac{dB}{dt} & = & - \frac{\rho}{k} (1-A-B)B. \label{damageds}
\end{eqnarray}
\end{subequations}
were now both $A(t), B(t)$ are positive functions depending only on time and describing the evolution of both tumor cell populations in systems without spatial inhomogeneities. The effect of radiotherapy given at times ($t_1,...,t_n)$ with doses $(d_1,...,d_n)$ and survival fractions $(S_f(d_1),...,S_f(d_n))$ in this simplified model is given by the equations

\begin{subequations}
\begin{eqnarray}
A(t^+_j) & = &  S_f(d_j) A(t_j^-),\\
B(t_j^+) &=& B(t_j^-) + \left[1-S_f(d_j)\right] A(t_j^-).
\end{eqnarray}
\end{subequations}

\subsection{Analysis of Eqs. \eqref{modeloEDOs}}

Since Eqs. (\ref{modeloEDOs}) correspond to an autonomous planar dynamical system, the possible dynamics in the phase space can be completely understood. First of all notice that 
there are two families of equilibria. First, the equilibrium point with $(A,B) = (0,0)$ and then the line of points $\mathcal{R}$ satisfying $A+B = 1$, with $A,B>0$. The first one is a saddle point, thus unstable and means that tumor cells tend to regrow no matter how small is their density. As to those in $\mathcal{R}=\{(a,1-a),\ 0<a<1\}$ the Jacobian matrix reads
\begin{equation}
J(\mathcal{R}) = \rho \begin{pmatrix} -a & -a \\ (1-a)/k & (1-a)/k \end{pmatrix} 
\end{equation}
 The eigenvalues of this Jacobian matrix are given by
 $$
 \lambda_{1}=0,\quad \lambda_{2}=\frac{\rho}{k}(1-a)-\rho a,
 $$
and the corresponding eigenvectors are $(1,-1)$ and $(1,-(1-a)/ka)$, respectively (for $\lambda_{2}\neq 0$). The equilibrium points on $\mathcal{R}$ are nonhyperbolic points. If $a<1/(k+1)$ then the fixed point $(a,1-a)$ possesses a local unstable manifold and a local center manifold. Otherwise, $(a,1-a)$ has a local stable manifold and a local center manifold. Thus, to get a heteroclinic orbit joining two points, say $(a_1,1-a_1)$ and $(a_2,1-a_2)$, with $a_2>a_1$, it is a necessary condition that $a_{2}>1/(k+1)$ and $a_{1}<1/(k+1)$. 

A straightforward application of the center manifold theory shows that the center manifold of $\mathcal{R}$ is the same set, i.e., $W^{c}(\mathcal{R})=\mathcal{R}$. Moreover, the points of $\mathcal{R}$ satisfying $a<1/(k+1)$ are unstable, while the points over the center manifold  satisfying $a>1/(k+1)$ are stable. 

\begin{center}
\begin{figure}
\begin{center}
\epsfig{file=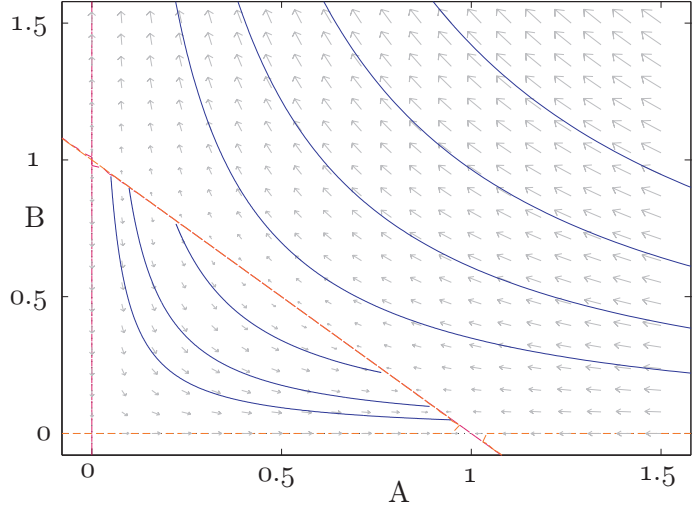,width=0.7\textwidth}
\end{center}
\caption{Phase portrait  of the dynamical system (\ref{modeloEDOs}). Shownare the velocity field (arrows) and some orbits (blue) including some heteroclinic orbits 
 connecting unstable equilibrium points on $\mathcal{R}$ with stable equilibria on the same set (blue). Also in red are shown the stable and unstable manifold of the point (0,0) (located on the axes), and the manifold of equilibria $\mathcal{R}$. The red lines determine the limits of the invariant region $\mathcal{S}$.
\label{mapa_fases}}
\end{figure}
\end{center}

The explicit form of the equation of the orbits can be obtained from Eqs. (\ref{modeloEDOs})
\begin{equation}
\frac{dA}{dB} = -k\frac{A}{B}
\end{equation}
what leads to 
\begin{equation}
A B^k = C,
\end{equation}
 with $C=A_{0}B_{0}^k$, where $A(0)=A_0>0, B(0)=B_0>0$. Thus, the orbits correspond to hyperbolas. Some orbits together with the velocity field are shown in Fig. \ref{mapa_fases}.  We want to note that the center manifold is given by the red line joining the points $(1,0)$ and $(0,1)$. 
The feasible region of our model is the set 
$$
\mathcal{S}=\{(A,B): A, B>0, \ A+B\leq 1\}.
$$
Since the set $\mathcal{S}$ is bounded by the center manifold and by the line orbits of the saddle point $A=0$ and $B=0$, it is straightforward to see that the region $\mathcal{S}$ is an invariant region, that is, all the orbits inside $\mathcal{S}$ belong to $\mathcal{S}$ for all times $t\in\mathbb{R}$. Therefore, all the orbits inside $\mathcal{S}$ start and end in the center manifold (except for the line orbits asymptotically approaching the saddle point $(0,0)$ for $t=\pm \infty$).

\subsection{Exact solutions}

It is interesting to note that in special cases it is possible to compute some explicit solutions for equations (\ref{modeloEDOs}). One of the most relevant cases corresponds to $k=1$, that, as has been discussed through the paper, is the biologically most relevant situation. 
In that case, substituting the orbits equation, $AB = C$ in (\ref{damageds}) we get
\begin{equation}\label{integral}
\int \frac{dB}{(B-\frac{1}{2})^2+C-\frac{1}{4}}=\rho(t-t_{0})
\end{equation}

It is interesting to note that for solutions starting in the feasible region $A+B \leq 1$ a simple calculation shows that $ 0<C < 1/4$. Let us define $Q_+ = \sqrt{\frac{1}{4}-C} < \frac{1}{2}$. In that case we can compute explicitly the integrals in Eq. (\ref{integral}) to get 
\begin{subequations}
\begin{eqnarray}
B(t) & = & \frac{1}{2} - Q_+ \tanh\left[Q_+ \rho(t-t_{0})\right] \\
A(t) & = & \frac{C}{\displaystyle{\frac{1}{2}-Q_+ \tanh\left[Q_+ \rho(t-t_{0})\right]}}
\end{eqnarray}
\end{subequations}

In the phase space, these solutions correspond to hyperbolas inside the region $\mathcal{S}$.

For the limit case  $C=1/4$, Eqs. (\ref{integral}) can be also solved explicitly to get
\begin{subequations}
\begin{eqnarray}
B(t) & = & \frac{1}{2}-\frac{1}{\rho(t-t_{0})}, \\
A(t) & = & \frac{1}{2-\displaystyle{\frac{4}{\rho(t-t_{0})}}}.
\end{eqnarray}
\end{subequations}

This is a special case, since that the solutions corresponds to hyperbolas through the point $(1/2,1/2)$, which is a point of the center manifold and for which $\lambda_{2}=0$. 

For completeness we also present the solutions for the case  $C>1/4$. In that case 
the solutions correspond to hyperbolas outside the region $\mathcal{S}$. Defining $Q_- = \sqrt{C-1/4}$ we get
\begin{subequations}
\begin{eqnarray}
B(t) & = & \frac{1}{2}+\ Q_- \tan\left[Q_- \rho(t-t_{0})\right], \\
A(t) & = & \frac{C}{\displaystyle{\frac{1}{2}+Q_- \tan\left[Q_- \rho(t-t_{0})\right]}}.
\end{eqnarray}
\end{subequations}


%

\end{document}